\documentstyle[preprint,aps,prd,eqsecnum,twoside,epsfig]{revtex}
\newcommand {\ve}{\varepsilon}
\newcommand {\pr}{\partial}
\newcommand {\cG}{\cal G}
\newcommand {\cD}{\cal D}
\newcommand {\bg}{\bar \gamma}
\newcommand {\G}{\Gamma}
\newcommand {\bp}{\bar \psi}

\newcommand {\vf}{\varphi}

\begin{document}
\baselineskip -24pt
\title{Spinor field with induced nonlinearity in a Bianchi type-I Universe} 
\author{Bijan Saha\\ 
Laboratory of Information Technologies\\ 
Joint Institute for Nuclear Research, Dubna\\ 
141980 Dubna, Moscow region, Russia\\ 
e-mail:  saha@thsun1.jinr.ru, bijan@cv.jinr.ru}
\maketitle 

\begin{abstract}
Self-consistent solutions to the spinor, scalar and BI gravitational 
field equations are obtained. The problems of initial singularity and 
asymptotically isotropization process of the initially anisotropic
space-time are studied. 
It is also shown that the introduction of the Cosmological constant
($\Lambda$-term) in the Lagrangian generates oscillations of the BI model,
which is not the case in absence of $\Lambda$ term. 
\end{abstract}


\section{Introduction}

The problem of initial singularity still remains at the center
of modern day cosmology. Though Big Bang theory is deep rooted 
among the scientists dealing with early day cosmology, it is 
natural to look back if one can model a Universe free from
initial singularity. 

The nonlinear generalization of classical field theory remains
one of the possible ways to over come the difficulties of the 
theory which considers elementary particles as mathematical 
points. The gravitational field equation is nonlinear by 
nature and the field itself is universal and unscreenable.
These properties lead to definite physical interest for the proper
gravitational field to be considered. 

FRW models are widely considered as good approximation of the present 
and early stages of the universe. But the large scale matter 
distribution in the observable universe, largely manifested in 
the form of discrete structures, does not exhibit homogeneity of 
a higher order. Recent space investigations detect anisotropy in 
the cosmic microwave background. 
The theoretical arguments~\cite{misner} and recent 
experimental datas that support the existence of an anisotropic 
phase that approaches an isotropic one leads to consider the
models of universe with anisotropic back-ground. Zel'dovich was 
first to assume that the early isotropization of cosmological 
expanding process can take place as a result of quantum effect of 
particle creation near singularity~\cite{zel1}. This assumption was 
further justified by several authors~\cite{lu1,lu2,hu1}.

A Bianchi type-I (BI) Universe, being the straightforward generalization 
of the flat Freidmann-Robertson-Walker (FRW) Universe, is one of the 
simplest models  
of an anisotropic Universe that describes a homogenous and spatially flat
Universe. Unlike the FRW Universe which has the same 
scale factor for each of the three spatial directions, a BI Universe
has a different scale factor in each direction, thereby introducing an
anisotropy to the system. It moreover has the agreeable property that
near the singularity it behaves like a Kasner Universe, even in the 
presence of matter, and consequently falls within the general analysis
of the singularity given by Belinskii et al~\cite{belinskii}. 
Also in a Universe filled with matter for $p\,=\,\zeta\,\ve, \quad 
\zeta < 1$, it has been shown that any initial anisotropy in a BI
Universe quickly dies away and a BI Universe eventually evolves
into a FRW Universe~\cite{jacobs}. Since the present-day Universe is 
surprisingly isotropic, this feature of the BI Universe makes it a prime 
candidate for studying the possible effects of an anisotropy in the early 
Universe on present-day observations. 

Nonlinear spinor field (NLSF) in external FRW cosmological
gravitational field was first studied by Shikin \cite{shikin}. 
Here we would like to note 
that the presence of a singular point to time in its space-time
metric is another important property of the isotropic model. The 
presence such a singular point means that the time is restricted.
The motivation for introduction a nonlinear term in the spinor field 
Lagrangian was to answer the natural question that arises in connection
with the presence of singular point, i.e., to what extent the  
presence of a singular point is an inherent property of the relativistic
cosmological models or whether it is only a consequence of specific
simplifying assumptions underlying these models. The study by Shikin
~\cite{shikin} shows that the presence of spinor filed nonlinearity 
is not enough to remove singularity in a FRW space-time. 
The natural choice was to introduce anisotropy in the model and analyze the 
nonlinear spinor field equations in an external BI universe that was 
carried out in~\cite{sahapfu1,sahactp1,sahajmp,sahaprd}
Recently, we study~\cite{pfu-l,sahal} the role of the cosmological 
constant ($\Lambda$) in the Lagrangian which together with 
Newton's gravitational constant ($G$) is considered as the fundamental
constants in Einstein's theory of gravity. We also considered an
interacting system of spinor and scalar field in a BI space-time
\cite{sahaizv,sahactp2,sahagrg}. 

In this report we consider a self-consistent system of spinor, scalar 
and Bianchi type-I gravitation fields in presence of perfect fluid and
$\Lambda$-term in an attepmt to \\
$\odot$
Obtain exact solutions to the nonlinear spinor, scalar and gravitational
fields;\\
$\odot$
Study initial singularity and possibilities of its elimination;\\
$\odot$
Isotropization of initially anisotropic cosmological model;\\
$\odot$
Possibility of the existence of oscillating models;\\
$\odot$
Role of $\Lambda$-term and perfect fluid in the evolution of BI universe.


\section{Fundamental Equations and general solutions}

The action of the nonlinear spinor, scalar and gravitational fields
can be written as
\begin{equation}
{\cal S}(g; \psi, \bp, \vf) = \int\, L \sqrt{-g} d\Omega
\label{action}
\end{equation}
with 
\begin{equation} 
L= L_{\rm g} + L_{\rm sp} + L_{\rm sc} + L_{\rm int} + L_{\rm m}.
\label{lag} 
\end{equation} 

Here $L_{\rm sp}$ is the spinor field Lagrangian given by
\begin{equation}
L_{\rm sp} = \frac{i}{2} 
\biggl[\bp \gamma^{\mu} \nabla_{\mu} \psi- \nabla_{\mu} \bar 
\psi \gamma^{\mu} \psi \biggr] - m\bp \psi.
\label{lspin}
\end{equation}
The mass-less scalar field Lagrangian is chosen as 
\begin{equation}
L_{\rm sc} = \frac{1}{2} \vf_{,\alpha}\vf^{,\alpha} 
\label{lsc}
\end{equation} 
while the interaction Lagrangian we choose~\cite{sahactp1,sahaizv,sahagrg}
\begin{equation}
L_{\rm int}= \frac{\lambda}{2}\,\vf_{,\alpha}\vf^{,\alpha} F,
\label{lint}
\end{equation}
with $\lambda$ being the coupling constant and $F$ is some 
arbitrary functions of invariants generated from the real bilinear 
forms of a spinor field. We choose the nonlinear  $F$ to be the 
function of $I =S^2 = (\bp \psi)^2$ and 
$J = P^2 = (i \bp \gamma^5 \psi)^2$, i.e., $F = F(I, J)$, 
that describes the nonlinearity in the most general of its 
form~\cite{sahaprd}. 

$L_{\rm g}$ corresponds to the gravitational field 
\begin{equation}
L_{\rm g} = \frac{R + 2 \Lambda}{2\kappa},
\label{lgrav}
\end{equation}
where $R$ is the scalar curvature, $\kappa = 8 \pi G$ with G 
being the Einstein's gravitational constant and $\Lambda$ is 
the cosmological constant. The gravitational field in our
case is given by a Bianchi type I (BI) metric 
in the form 
\begin{equation} 
ds^2 = dt^2 - a^2(t) dx^2 - b^2(t) dy^2 - c^2(t) dz^2, 
\label{BI1}
\end{equation}
with $a(t), b(t), c(t)$ being the functions of time only.

Variation of (\ref{action}) with respect to spinor field $\psi\,(\bp)$
gives nonlinear spinor field equations
\begin{mathletters}
\label{speq}
\begin{eqnarray}
i\gamma^\mu \nabla_\mu \psi - m \psi + {\cD} \psi + 
{\cG} i \gamma^5 \psi &=&0, \label{speq1} \\
i \nabla_\mu \bp \gamma^\mu +  m \bp - {\cD} \bp - 
{\cG} i \bp \gamma^5 &=& 0, \label{speq2}
\end{eqnarray}
\end{mathletters}
where we denote
$$ {\cD} =  \lambda S \vf_{,\alpha}\vf^{,\alpha} \frac{\pr F}{\pr I}, 
\quad
{\cG} =  \lambda P \vf_{,\alpha}\vf^{,\alpha} \frac{\pr F}{\pr J},$$
whereas, variation of (\ref{action}) with respect to scalar field
yields the following scalar field equation
\begin{equation}
\frac{1}{\sqrt{-g}} \frac{\pr}{\pr x^\nu} \Bigl(\sqrt{-g} g^{\nu\mu}
(1 + \lambda F) \vf_{,\mu}\Bigr) 
= 0. \label{scfe}
\end{equation}

Varying (\ref{action}) with respect to metric tensor $g_{\mu\nu}$ 
one finds the Einstein's field equation
\begin{mathletters}
\label{BID}
\begin{eqnarray}
\frac{\ddot b}{b} +\frac{\ddot c}{c} + \frac{\dot b}{b}\frac{\dot 
c}{c}&=&  \kappa T_{1}^{1} -\Lambda,\label{11}\\
\frac{\ddot c}{c} +\frac{\ddot a}{a} + \frac{\dot c}{c}\frac{\dot 
a}{a}&=&  \kappa T_{2}^{2} - \Lambda,\label{22}\\
\frac{\ddot a}{a} +\frac{\ddot b}{b} + \frac{\dot a}{a}\frac{\dot 
b}{b}&=&  \kappa T_{3}^{3} - \Lambda,\label{33}\\
\frac{\dot a}{a}\frac{\dot b}{b} +\frac{\dot b}{b}\frac{\dot c}{c} 
+\frac{\dot c}{c}\frac{\dot a}{a}&=&  \kappa T_{0}^{0} - \Lambda,
\label{00}
\end{eqnarray}
\end{mathletters}
where point means differentiation with respect to $t$ 
and $T_{\nu}^{\mu}$ is the energy-momentum tensor
of the material field given by
\begin{equation}
T_{\mu}^{\nu} = T_{{\rm sp}\,\mu}^{\,\,\,\nu} + T_{{\rm sc}\,\mu}^{\,\,\,\nu}
+ T_{{\rm int}\,\mu}^{\,\,\,\nu} + T_{{\rm m}\,\mu}^{\,\,\,\nu}.
\label{tem}
\end{equation}
Here $T_{{\rm sp}\,\mu}^{\,\,\,\nu}$ is the energy-momentum tensor of 
the spinor field  
\begin{equation}
T_{{\rm sp}\,\mu}^{\,\,\,\rho}=\frac{i}{4} g^{\rho\nu} \biggl(\bp \gamma_\mu 
\nabla_\nu \psi + \bp \gamma_\nu \nabla_\mu \psi - \nabla_\mu \bar 
\psi \gamma_\nu \psi - \nabla_\nu \bp \gamma_\mu \psi \biggr) \,-
\delta_{\mu}^{\rho}L_{sp}
\label{temsp}
\end{equation}
where $L_{sp}$ with respect to (\ref{speq}) takes the form
\begin{equation}
L_{sp} = -\bigl({\cD} S + {\cG} P\bigr).
\label{lsp}
\end{equation}
The energy-momentum tensor of the scalar field 
$T_{{\rm sc}\,\mu}^{\,\,\,\nu}$ is given by 
\begin{equation}
T_{{\rm sc}\,\mu}^{\,\,\,\nu}= \vf_{,\mu}\vf^{,\nu}
- \delta_{\mu}^{\nu} L_{\rm sc}.
\label{temsc}
\end{equation}
For the interaction field we find
\begin{equation}
T_{{\rm int}\,\mu}^{\,\,\,\nu}= \lambda F \vf_{,\mu}\vf^{,\nu}
 -  \delta_{\mu}^{\nu} L_{\rm int}.
\label{temint}
\end{equation}
$T_{\mu\,(m)}^{\nu} = (\ve,\,-p,\,-p,\,-p)$
is the energy-momentum tensor of a perfect fluid. 
Energy $\ve$ is related to the pressure $p$ by the equation 
of state $p\,=\,\zeta\,\ve$.  Here $\zeta$ varies between the
interval $0\,\le\, \zeta\,\le\,1$, whereas $\zeta\,=\,0$ describes
the dust Universe, $\zeta\,=\,\frac{1}{3}$ presents radiation Universe,
$\frac{1}{3}\,<\,\zeta\,<\,1$ ascribes hard Universe and $\zeta\,=\,1$
corresponds to the stiff matter. 

The explicit form of the covariant derivative
of spinor is ~\cite{zhelnorovich,brill}
\begin{mathletters}
\label{cvd}
\begin{eqnarray} 
\nabla_\mu \psi &=& \frac{\partial \psi}{\partial x^\mu} -\G_\mu \psi, \\
\nabla_\mu \bp &=& \frac{\partial \bp}{\partial x^\mu} + \bp \G_\mu, 
\end{eqnarray} 
\end{mathletters}
where $\G_\mu(x)$ are spinor affine connection matrices.  
For the metric (\ref{BI1}) one has the following components
of the affine spinor connections 
\begin{eqnarray} 
\G_0 = 0, \quad 
\G_1 = \frac{1}{2}\dot a(t) \bg^1 \bg^0, \quad
\G_2 = \frac{1}{2}\dot b(t) \bg^2 \bg^0, \quad 
\G_3 = \frac{1}{2}\dot c(t) \bg^3 \bg^0. 
\end{eqnarray}
 
We study the space-independent solutions to the spinor 
and scalar field equations (\ref{speq}) and (\ref{scfe}) so that 
$\psi=\psi(t)$ and $\vf = \vf(t)$.
defining
\begin{equation}
\tau = a b c = \sqrt{-g}
\label{taudef}
\end{equation}
from (\ref{scfe}) for the scalar field  we have
\begin{equation}
\vf = C \int \frac{dt}{\tau (1 + \lambda F)}.
\label{sfsol}
\end{equation}

The spinor field equation (\ref{speq1})  we rewrite as
\begin{equation} i\bg^0 
\biggl(\frac{\partial}{\partial t} +\frac{\dot \tau}{2 \tau} \biggr) \psi 
- m \psi + {\cD}\psi + {\cG} i \gamma^5 \psi = 0.
\label{spq}
\end{equation} 
Setting $V_j(t) = \sqrt{\tau} \psi_j(t), \quad j=1,2,3,4,$ from 
(\ref{spq}) one deduces the following system of equations:  
\begin{mathletters}
\label{V}
\begin{eqnarray} 
\dot{V}_{1} + i (m - {\cD}) V_{1} - {\cG} V_{3} &=& 0, \\
\dot{V}_{2} + i (m - {\cD}) V_{2} - {\cG} V_{4} &=& 0, \\
\dot{V}_{3} - i (m - {\cD}) V_{3} + {\cG} V_{1} &=& 0, \\
\dot{V}_{4} - i (m - {\cD}) V_{4} + {\cG} V_{2} &=& 0. 
\end{eqnarray} 
\end{mathletters}

From (\ref{speq1}) we also write the equations for the invariants
$S,\quad P$ and $A = \bp \bg^5 \bg^0 \psi$
\begin{mathletters}
\begin{eqnarray}
{\dot S_0} - 2 {\cG}\, A_0 &=& 0, \label{S0}\\
{\dot P_0} - 2 (m - {\cD})\, A_0 &=& 0, \label{P0}\\
{\dot A_0} + 2 (m - {\cD})\, P_0 + 2 {\cG} S_0 &=& 0, \label{A0} 
\end{eqnarray}
\end{mathletters}
where $S_0 = \tau S, \quad P_0 = \tau P$, and $ A_0 = \tau A$,
leading to the following relation
\begin{equation}
S^2 + P^2 + A^2 =  C^2/ \tau^2, \qquad C^2 = {\rm const.}
\label{inv1}
\end{equation}

Giving the concrete form of $F$ from (\ref{V}) one writes
the spinor functions in explicit form and
using the solutions obtained one can write the components of
spinor current:
\begin{equation}
j^\mu = \bp \gamma^\mu \psi.
\label{spincur}
\end{equation}
The component $j^0$ 
\begin{equation}
j^0 = \frac{1}{\tau}
\bigl[V_{1}^{*} V_{1} + V_{2}^{*} V_{2} + V_{3}^{*} V_{3}
+ V_{4}^{*} V_{4}\bigr], 
\end{equation}
defines the charge density of spinor field 
that has the following chronometric-invariant form 
\begin{equation}
\varrho = (j_0\cdot j^0)^{1/2}. 
\label{rho}
\end{equation}
The total charge of spinor field is defined as
\begin{equation}
Q = \int\limits_{-\infty}^{\infty} \varrho \sqrt{-^3 g} dx dy dz =
   \varrho \tau V,
\label{charge}
\end{equation}
where $V$ is the volume. From the spin tensor
\begin{equation}
S^{\mu\nu,\epsilon} = \frac{1}{4}\bp \bigl\{\gamma^\epsilon
\sigma^{\mu\nu}+\sigma^{\mu\nu}\gamma^\epsilon\bigr\} \psi.
\label{spin}
\end{equation}
one finds chronometric invariant spin tensor 
\begin{equation}
S_{{\rm ch}}^{ij,0} = \bigl(S_{ij,0} S^{ij,0}\bigr)^{1/2},
\label{chij}
\end{equation} 
and the projection of the spin vector on $k$ axis 
\begin{equation}
S_k = \int\limits_{-\infty}^{\infty} S_{{\rm ch}}^{ij,0} 
\sqrt{-^3 g} dx dy dz = S_{{\rm ch}}^{ij,0} \tau V. 
\label{proj}
\end{equation} 

Let us now solve the Einstein equations. To do it we first write the 
expressions for the components of the energy-momentum tensor explicitly:
\begin{eqnarray}
\label{total}
T_{0}^{0} &=& mS + \frac{C^2}{2\tau^2 (1+\lambda F)} + \ve, \nonumber\\
\\
T_{1}^{1} &=& T_{2}^{2} = T_{3}^{3} =
{\cD} S + {\cG} P - \frac{C^2}{2\tau^2 (1+\lambda F)} 
- p.\nonumber 
\end{eqnarray}
In account of (\ref{total}) from  (\ref{11}), (\ref{22}),(\ref{33})
we find the metric functions~\cite{sahaprd}
\begin{eqnarray} 
a(t) &=& 
(D_{1}^{2}D_{3})^{1/3}\tau^{1/3}\mbox{exp}\biggl[\frac{2 X_1 + X_3 
}{3} \int\,\frac{dt}{\tau (t)} \biggr], \label{a} \\
b(t) &=& 
(D_{1}^{-1}D_{3})^{1/3}\tau^{1/3}\mbox{exp}\biggl[-\frac{X_1 - X_3 
}{3} \int\,\frac{dt}{\tau (t)} \biggr], \label{b} \\
c(t) &=& 
(D_{1}D_{3}^{2})^{-1/3}\tau^{1/3}\mbox{exp}\biggl[-\frac{X_1 + 2 X_3 
}{3} \int\,\frac{dt}{\tau (t)} \biggr].  \label{c}
\end{eqnarray}
As one sees from (\ref{a}), (\ref{b}) and   (\ref{c}), for $\tau = t^n$
with $n > 1$ the exponent tends to unity at large $t$, and the 
anisotropic model becomes isotropic one. 
Let us also write the invariants of gravitational field. They are
the Ricci scalar $I_1 = R \approx 1/\tau$,
$I_2 = R_{\mu\nu}R^{\mu\nu} \equiv R_{\mu}^{\nu} R_{\nu}^{\mu}
\approx 1/\tau^3$ and the Kretschmann scalar
$I_3 = R_{\alpha\beta\mu\nu}R^{\alpha\beta\mu\nu} \approx 1/\tau^6$.
As we see, the space-time becomes singular at a point where $\tau = 0$,
as well as the scalar and spinor fields. Thus we see, all the functions
in question are expressed via $\tau$. In what follows, we write the
equation for $\tau$ and study it in details.  

Summation of Einstein equations (\ref{11}), (\ref{22}),(\ref{33}) and 
(\ref{00}) multiplied by 3 gives
\begin{equation}
\frac{\ddot 
\tau}{\tau}= \frac{3}{2}\kappa \Bigl(mS + {\cD} S + {\cG} P + \ve -p
\Bigr) - 3 \Lambda. 
\label{dtau1}
\end{equation} 
For the right-hand-side of (\ref{dtau1}) to be a function
of $\tau$ only, the solution to this equation is well-known~\cite{kamke}.

From energy-momentum conservation law 
\begin{equation}
T_{\mu;\nu}^{\nu} = T_{\mu,\nu}^{\nu} + \G_{\rho\nu}^{\nu} T_{\mu}^{\rho}
- \G_{\mu\nu}^{\rho} T_{\rho}^{\nu} = 0
\end{equation}
in account of $p = \zeta \ve$ we obtain 
\begin{equation}
\ve = \frac{\ve_0}{\tau^{1+\zeta}},\quad 
p = \frac{\zeta \ve_0}{\tau^{1+\zeta}}.
\label{vep}
\end{equation}
In our consideration  of $F$ as a function of $I$, $J$ or $I\pm J$
we get these arguments, as well as ${\cD}$ and ${\cG}$ as functions of
$\tau$. Hence the right-hand-side of (\ref{dtau1}) is a function
of $\tau$ only. 
Then (\ref{dtau1}), multiplied by $2 {\dot \tau}$ can be written as
\begin{equation}
2 {\dot \tau}\,{\ddot \tau} = \bigl[3\bigl(\kappa 
(mS + {\cD} S + {\cG} P + \ve - p)
- 2 \Lambda \bigr) \tau\bigr] {\dot \tau} = \Psi(\tau) {\dot \tau}  
\label{taug}
\end{equation} 
Solution to the equation (\ref{taug}) we write in quadrature
\begin{equation}
\int\,\frac{d \tau}{\sqrt{\int \Psi (\tau) d \tau}} = t + t_0.
\label{quad}
\end{equation}
From here on we set $t_0 = 0$, as it only gives the shift of the initial time. 
Given the explicit form of $F(I,\,J)$, from (\ref{quad}) one finds 
concrete function $\tau(t)$. Once the value of $\tau$ is obtained, one can 
get expressions for components $\psi_j(t), \quad j = 1, 2, 3, 4.$
Thus the initial systems of spinor, scalar and gravitational field 
equations are completely integrated.


\section{Analysis of the results} 

In this section we analyze the system of spinor, scalar and
gravitational field for some concrete form of $F$.

As it was mentioned earlier, we choose $F$ to be a function of $I$ and $J$
only. We first consider the case with $F=F(I)$, that will followed by
$F=F(J)$. 

{\bf case (i)} Let us consider the case with $F=F(I)$. In this case
from (\ref{S0}) one finds
\begin{equation}
S = C_0/\tau, \quad C_0 = {\rm const.}
\label{stau}
\end{equation}
In this case for spinor field components we find~\cite{sahaprd}
\begin{eqnarray} 
\psi_1(t) &=& \frac{C_1}{\sqrt{\tau}} e^{-i\beta}, \quad
\psi_2(t) = \frac{C_2}{\sqrt{\tau}} e^{-i\beta},  \nonumber\\
\label{spef}\\
\psi_3(t) &=& \frac{C_3}{\sqrt{\tau}} e^{i\beta}, \quad
\psi_4(t) = \frac{C_4}{\sqrt{\tau}} e^{i\beta},
\nonumber
\end{eqnarray} 
with $C_i$ being the integration constants and
are related to $C_0$ as 
$C_0 = C_{1}^{2} + C_{2}^{2} - C_{3}^{2} - C_{4}^{2}.$ Here
$\beta = \int(m - {\cD}) dt$.

For the components of the spin current from (\ref{spincur}) we find
\begin{eqnarray}
j^0 &=& \frac{1}{\tau}
\bigl[C_{1}^{2} + C_{2}^{2} + C_{3}^{2} + C_{4}^{2}\bigr],\quad
j^1 = \frac{2}{a\tau}
\bigl[C_{1} C_{4} + C_{2} C_{3}\bigr] {\rm cos}(2\beta),
\nonumber \\
j^2 &=& \frac{2}{b\tau}
\bigl[C_{1} C_{4} - C_{2} C_{3}\bigr] {\rm sin}(2\beta),\quad
j^3 = \frac{2}{c\tau}
\bigl[C_{1} C_{3} - C_{2} C_{4}\bigr] {\rm cos}(2\beta), \nonumber
\end{eqnarray}
whereas, for the projection of spin vectors on the $X$, $Y$ and $Z$
axis we find
\begin{eqnarray}
S^{23,0} = \frac{C_1 C_2 + C_3 C_4}{b c\tau},\quad
S^{31,0} = 0,\quad
S^{12,0} = \frac{C_1^2 - C_2^2 + C_3^2 - C_4^2}{2ab\tau}. \nonumber
\end{eqnarray}
Total charge of the system in a volume $\cal{V}$ in this case is
\begin{equation}
Q = [C_1^2 + C_{2}^{2} + C_{3}^{2} + C_{4}^{2}] \cal{V}.
\end{equation}
Thus, for $\tau \ne 0$ the components of spin current and
the projection of spin vectors are singularity-free and the total 
charge of the system in a finite volume is always finite.

Let us set $F = S^n$. In this case for $\tau$ we write the second
order equation 
\begin{equation}
\ddot{\tau} = \frac{3\kappa}{2}\Bigl(m +\lambda n  
\tau^{n-1}/2
(\lambda  + \tau^n)^2+ \ve_0 (1-\zeta)/\tau^\zeta\Bigr)-3 \Lambda \tau,
\label{nueq}
\end{equation}
or the one of first order
\begin{equation}
\frac{d \tau}{\sqrt{\kappa[m \tau + \lambda  \tau^n/2(\lambda  
+\tau^n)] -\Lambda \tau^2 +\ve_0 \tau^{1-\zeta} +y_1^2}} = \sqrt{3} t.
\label{quads}
\end{equation}
Here we set $C=1$ and $C_0=1$. It should be emphasized that, 
$\tau$ cannot be negative, i.e., it has a lower limit. 
On the other hand, for $\Lambda > 0$, for the 
integrant of (\ref{quads}) to be positive definite, $\tau$
should have some upper limit. $y_1$ in (\ref{quads}) is an arbitrary
integration constant. Giving boundary conditions the equation
(\ref{nueq}) can be solved by continuous analog of Newton
method~\cite{zhidkov}, whereas given initial (or asymptotical) value
one can numerically solve (\ref{quads}) by Runge-Kutta method.
Since $y_1$ in (\ref{quads}) is arbitrary, we use continuous analog
of Newton method.

Let us formulate the boundary condition. At initial stage 
perfect fluid plays the principal role and $\tau$ at this stage 
obeys
\begin{equation}
\ddot{\tau} = \frac{3\kappa}{2} \ve_0 (1-\zeta)/\tau^\zeta,
\end{equation}
with the solution
\begin{equation}
\tau |_{t \to 0} = A = \Bigl[\frac{1+\zeta}{2}(\sqrt{3 \kappa \ve_0} t + c_1)
\Bigr]^{2/(1+\zeta)}, \quad  c_1={\rm const.}
\label{lb}
\end{equation}
whereas, at large $t$ we have for $\tau$
\begin{equation}
\ddot{\tau} = \frac{3\kappa}{2} m - 3 \Lambda \tau,
\end{equation}
with the solutions
\begin{equation}
\tau \Bigl|_{t \to \infty} = B = \left\{\begin{array}{cc}
(3/4) \kappa m  t^2,  & \Lambda = 0, \\ 
{\rm sin}(\sqrt{3 \Lambda}\, t) + (\kappa m /2 \Lambda), & \Lambda > 0,\\
{\rm sinh}(\sqrt{-3 \Lambda}\, t) + (\kappa m /2 \Lambda), & \Lambda < 0.
\end{array}\right.
\label{rb}
\end{equation}
Note that, for the solutions to be regular, $\tau$ should be non-zero.
On the other hand, being  $\sqrt{-g}$, $\tau$ is non-negative. As one sees,
in this case the right hand side of (\ref{nueq}) is a continuous function.
A positive $c_1$ in (\ref{lb}) provides us non-zero $\tau$ as $t \to 0$,
whereas, manipulating the parameters $\kappa, m, \Lambda$ in (\ref{rb})
we always have a positive $\tau$ as $t \to \infty.$

Further, setting 
\begin{equation}
\tau = \xi + [(B-A)/(t_N-t_0)] t + (t_N A - t_0 B)/(t_N-t_0),
\label{subst}
\end{equation}
into (\ref{nueq}) we get the 
second order equation for $\xi (t)$ 
\begin{equation}
\ddot{\xi} = \Phi(\xi),
\label{xi}
\end{equation}
where $Phi(\xi)$ coincides with the right hand side of (\ref{nueq})
with $\tau$ substituted by (\ref{subst}). The boundary conditions
in this case regardless to the value of $\Lambda$ read
\begin{eqnarray}
\xi (t) \bigl|_{t = t_0} = 0,  \quad
\xi (t) \bigl|_{t = t_N} = 0.  
\label{bc}
\end{eqnarray}
Let us now solve (\ref{xi}) together with (\ref{bc}). For this we rewrite
(\ref{xi}) in the form
\begin{equation}
\Psi (\xi) = \frac{d^2 \xi}{d t^2} - \Phi (\xi) = 0.
\label{Psi}
\end{equation}
Assume
\begin{equation}
\xi = \xi (t, \sigma), \quad 0 < \sigma < +\infty,
\label{param}
\end{equation}
with $\sigma$ being some parameter. Further we demand
\begin{equation}
\frac{d \Psi [\xi(t,\sigma)]}{d \sigma} = - \Psi[\xi(t,\sigma)].
\label{dem}
\end{equation}
In this case we obtain 
\begin{equation}
\Psi [\xi(t,\sigma)] = \Psi[\xi(t,0)] e^{-\sigma}
\Biggl|_{\sigma \to +\infty} \to 0.
\end{equation}
It means for $\sigma$ large enough we tend to our initial problem.
Let us assume that $\xi$ possesses continuous
differentiation. Then the left hand side of (\ref{dem}) can be written as
\begin{equation}
\frac{\partial^2}{\partial t^2}\Bigl(\frac{d \xi}{d \sigma}\Bigr)
- \frac{\partial \Phi}{\partial \xi} \frac{d \xi}{d \sigma}
=\frac{d^2 \eta}{d t^2} - \frac{\partial \Phi}{\partial \xi} \eta
\end{equation} 
where we define
\begin{equation}
\frac{d \xi}{d \sigma} = \eta.
\label{etadef}
\end{equation}
Thus we now have the following equation for $\eta$:
\begin{equation}
\frac{d^2 \eta}{d t^2} - \frac{\partial \Phi}{\partial \xi} \eta
= - \bigl[\frac{d^2 \xi}{d t^2} - \Phi (\xi)\bigr],
\label{eta}
\end{equation} 
with the boundary conditions
\begin{equation}
\eta (0,\sigma) = 0, \quad \eta (\infty, \sigma) = 0.
\label{etabc}
\end{equation}
Thus we now have the boundary value problem (\ref{eta}) and (\ref{etabc}).
To solve (\ref{eta}) we need to give some initial value of $\xi$.
Setting
\begin{equation}
\xi (t, \sigma=0) = \xi_0,
\end{equation}
from (\ref{eta}) we find $\eta (t, 0)$. Further from (\ref{etadef}) we find
\begin{equation}
\xi (t,\sigma) = \xi_0 + (\Delta\sigma) \eta(t, 0),
\label{1step}
\end{equation}
where $\Delta\sigma$ is the step along $\sigma$ and lies between $[0,1]$,
i.e., $0 \le \Delta\sigma \le 1$. For $\Delta\sigma = 1$ we have the 
classical Newton method.
This process is performed again and again until $\eta$ becomes small enough.

Function $\xi_0$ is completely problem depended. Since $\xi = 0$ at both
left and right boundary, we have to choose $\xi_0$ that is zero at the
end-points. In our particular case,
as an initial approximation of $\xi$ one may consider any non-negative
function that tends to zero at $t_0$ ($t = 0$) and $t_N$ ($t = \infty$),
e.g., 
\begin{equation}
\xi(t) = (t-t_0)\times(t_N-t),
\label{ini1}
\end{equation}
or,
\begin{equation}
\xi(t) = {\rm sin}(\frac{\pi}{t_N-t_0} t).  
\label{ini2}
\end{equation}
Note that the equation (\ref{nueq}) with boundary conditions
(\ref{lb}) and (\ref{rb}) is a multi-parametric problem. The parameters
in questions are the Einstein's gravitational constant $\kappa$, 
spinor mass $m$, coupling constant $\lambda$,
power of nonlinearity $n$, $\zeta$, cosmological constant $\Lambda$
and the constant of integration $c_1$. We choose $\kappa = 1$ and
$c_1 = 1.0E-10$ for all cases. It should be emphasized that the
power of nonlinearity $n$ may be any number including a negative one. 
In our case we set $n=2$. Note that $\tau$ bahaves more or less in the 
same manner when $n$ is even, whereas, for odd $n$ we have totally
different picture depending on its concrete value.
For spinor mass we put $m=1$. The coupling
constant $\lambda$ may be both positive and negative and plays a 
crucial role here. We set $\lambda = 1$ and $\lambda = 0$ for interacting
and case with minimal coupling, respectively. The parameter $\zeta$ 
describes perfect fluid and takes the values $\zeta=0$ (dust),
$\zeta=1/3$ (radiation), $\zeta = 2/3$ (hard universe) and $\zeta=1$
(stiff matter). In the figures the notations "d", "r", "h" and "s" stand for
dust, radiation, hard universe and stiff matter, respectively. Finally,
for cosmological constant we set $\Lambda = (-0.4, 0.0, 0.4)$. Changing
these parameters we can manipulate the evolution of $\tau$, e.g.,
for $\Lambda>0$ we can obtain $\tau$ that is positive everywhere. The 
numerical experiment with different values of these parameters we leave
for some future papers. Here we solve the problem for some definite values
mentioned earlier and chalk out the main feature. For a positive $\Lambda$
we always have oscillatory mode, whereas for negative $\Lambda$ we have
inflation-like solution. In absence of $\Lambda$-term we have the solution
that steadily approaches to isotropic stage for a massive spinor, while
as will be shown later in case of $F=P^n$, for a massless spinor field
isotropization does not take place. It should be mentioned that interaction
term plays no significant role in this case, hence we have the identical
picture for minimal coupling and direct interaction between spinor and
scalar fields. Spinor mass plays a crucial role in absence of 
$\Lambda$-term. Note that, for a negative $\Lambda$-term
evolution of $\tau$ is completely dominated by $\Lambda$, and independent
to the value of $m$, $n$ and $\zeta$ we always have the same picture. As it
was mentioned earlier, at a space-time point where $\tau=0$, we have 
singularity, whereas, in the intervals where $\tau < 0$, corresponding
solution is not a physical one. We would like to remark that, giving suitable
values for the parameters, it is possible to find solutions those are
everywhere non-negative, but singularity-free solutions can be achieved
only for $\Lambda>0$. It should also be emphasized that, regular solutions 
may be obtained by virtue of spinor field nonlinearity~\cite{sahaprd} or
interaction term~\cite{sahagrg}, but this singularity-free solutions 
are attained at the cost of broken dominant energy condition in 
Hawking-Penrose theorem~\cite{zeldovich,hawking}.

The energy density of the system in this case has the form
\begin{equation}
T_{0}^{0} = \frac{m C_0}{\tau} +\frac{C^2 \tau^{n-2}}
{2(\tau^n +\lambda C_0^n)} +\frac{\ve_0}{\tau^{1+\zeta}}.
\label{ens}
\end{equation} 
For the total energy we then have
\begin{equation}
E = \int\limits_{-\infty}^{\infty} T_{0}^{0} \sqrt{-^3 g} dx dy dz
=\Bigl[m C_0 + \frac{C^2 \tau^{n-1}}
{2(\tau^n +\lambda C_0^n)} +\frac{\ve_0}{\tau^{\zeta}}\Bigr] \cal{V}.
\label{toten}
\end{equation}
As one sees from (\ref{toten}), if $\tau$ is nontrivial, then the total
energy of the system in a finite volume $\cal{V}$ is always finite.

{\bf case (ii)} Let us consider the case when $F=F(J)$. 
In the case considered we assume the spinor field to be massless.
It gives ${\cD} = 0$. Let us note that, in the unified 
nonlinear spinor theory of Heisenberg, the massive term remains 
absent, and according to Heisenberg, the particle mass should be 
obtained as a result of quantization of spinor prematter~
\cite{massless}. In the nonlinear generalization of classical field 
equations, the massive term does not possess the significance that 
it possesses in the linear one, as it by no means defines total 
energy (or mass) of the nonlinear field system. Thus without losing 
the generality we can consider mass-less spinor field putting $m\,=\,0.$ 
Then from (\ref{P0}) one gets
\begin{equation}
P = D_0/\tau, \quad D_0 = {\rm const.}
\end{equation}
In this case the spinor field components take the form
\begin{eqnarray}
\psi_1 &=&\frac{1}{\sqrt{\tau}} \bigl(D_1 e^{i \sigma} + 
iD_3 e^{-i\sigma}\bigr), \quad
\psi_2 =\frac{1}{\sqrt{\tau}} \bigl(D_2 e^{i \sigma} + 
iD_4 e^{-i\sigma}\bigr), \nonumber \\
\label{J}\\
\psi_3 &=&\frac{1}{\sqrt{\tau}} \bigl(iD_1 e^{i \sigma} + 
D_3 e^{-i \sigma}\bigr),\quad
\psi_4 =\frac{1}{\sqrt{\tau}} \bigl(iD_2 e^{i \sigma} + 
D_4 e^{-i\sigma}\bigr). \nonumber
\end{eqnarray} 
The integration constants $D_i$
are connected to $D_0$ by
$D_0=2\,(D_{1}^{2} + D_{2}^{2} - D_{3}^{2} -D_{4}^{2}).$
Here we set $\sigma = \int {\cG} dt$. 

For the components of the spin current from (\ref{spincur}) we find
\begin{eqnarray}
j^0 &=& \frac{2}{\tau}
\bigl[D_{1}^{2} + D_{2}^{2} + D_{3}^{2} + D_{4}^{2}\bigr],\quad
j^1 = \frac{4}{a\tau}
\bigl[D_{2} D_{3} + D_{1} D_{4}\bigr] {\rm cos}(2 \sigma), \nonumber\\
j^2 &=& \frac{4}{b\tau}
\bigl[D_{2} D_{3} - D_{1} D_{4}\bigr] {\rm sin}(2 \sigma),\quad
j^3 = \frac{4}{c\tau}
\bigl[D_{1} D_{3} - D_{2} D_{4}\bigr] {\rm cos}(2 \sigma), \nonumber
\end{eqnarray}
whereas, for the projection of spin vectors on the $X$, $Y$ and $Z$
axis we find
\begin{eqnarray}
S^{23,0} = \frac{2(D_{1} D_{2} + D_{3} D_{4})}{b c\tau},\quad
S^{31,0} = 0,\quad
S^{12,0} = \frac{D_{1}^{2} - D_{2}^{2} + D_{3}^{2} - D_{4}^{2}}{2ab\tau}
\nonumber
\end{eqnarray}
Let us now choose $F = P^n$. In this case in view of massless spinor
field for $\tau$ we obtain 
\begin{equation}
\ddot{\tau} = \frac{3\kappa}{2}\Bigl(\lambda n  
\tau^{n-1}/2
(\lambda  + \tau^n)^2+ \ve_0 (1-\zeta)/\tau^\zeta\Bigr)-3 \Lambda \tau.
\label{nueqp}
\end{equation}
At $t = 0$ we have the boundary condition  as in the previous case, i.e.,
\begin{equation}
\tau |_{t \to 0} = \Bigl[\frac{1+\zeta}{2}(\sqrt{3 \kappa \ve_0} t + c_1)
\Bigr]^{2/(1+\zeta)}, \quad  c_1={\rm const.},
\label{lbcp}
\end{equation}
whereas, at large $t$ we have 
\begin{equation}
\tau(t) \Bigl|_{t \to \infty} = \left\{\begin{array}{ccc}
q_1\,t + q_2,  & \Lambda = 0,\\
{\rm sin}(\sqrt{3 \Lambda}\,t), & \Lambda > 0 \\
{\rm sinh}(\sqrt{-3 \Lambda}\,t), & \Lambda < 0. 
\end{array}\right.
\label{rbcp}
\end{equation}
Here $q_1,\,q_2$ are constants of integration.
We solve the equations (\ref{nueqp}), (\ref{lbcp}) and (\ref{rbcp}) 
numerically using continuous analog of Newton method. 

It should be mentioned that the results for $\zeta=0$ and $\zeta=1$
are obtained with $|\eta| < 1.0E-13$. Iteration in this case 
varies from 20 to 80. For $\zeta = 1/3$ and $\zeta = 2/3$ the value 
of $|\eta|$ 
is rather rough. In case of massless spinor field within the scope
of parameters chosen here, no solutions were available with a positive 
$\Lambda$ for radiation and hard universe. Corresponding  
solutions are presented graphically in the figures followed.
As one observes in the following figures, solutions are singular
at initial time and isotropization process takes place for massive spinor 
field. For positive $\Lambda$-term solutions are always oscillatory
and the amplitude of oscillations decreases as the value of $\zeta$
increases. In absence of $\Lambda$-term beside the spinor mass perfect 
fluid plays a significant role. Especially, for dust and radiation,
expansion of $\tau$ starts beginning from some positive value of $t$.

\begin{figure}
\hspace{2cm}\epsfig{file=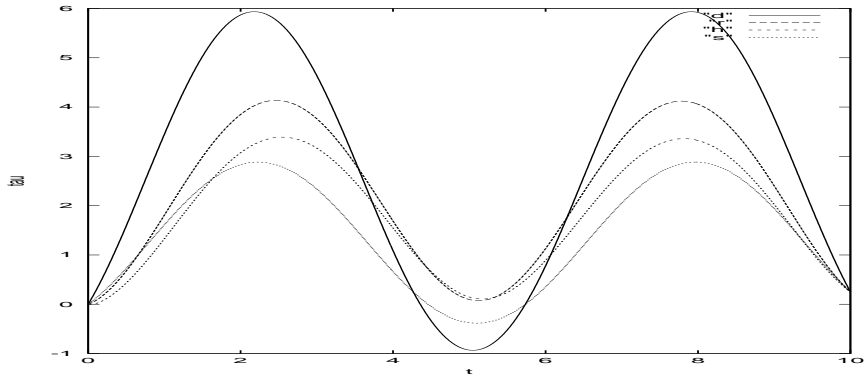,height=5cm,width=12cm,angle=0}
\vspace{.5cm}
\caption{Evolution of $\tau$ in case of minimal coupling with positive
$\Lambda$.}
\end{figure}

\begin{figure}
\hspace{2cm}\epsfig{file=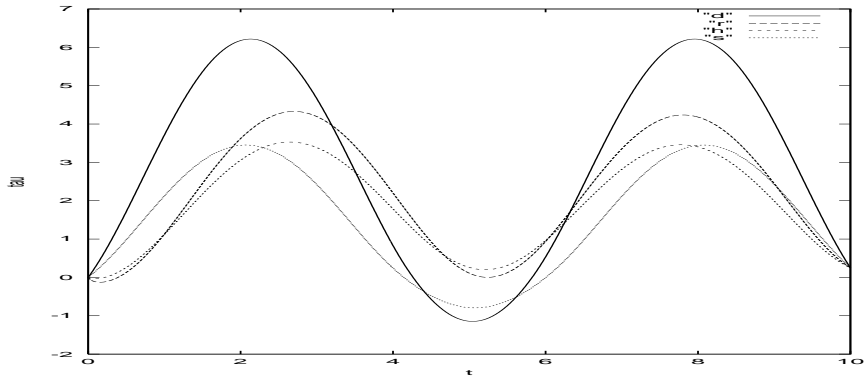,height=5cm,width=12cm,angle=0}
\vspace{.5cm}
\caption{Evolution of $\tau$ for a positive $\Lambda$ term with direct
interaction between spinor and scalar fields. As one sees, interacting
term is significant only at the vicinity of $t=0$.}
\end{figure}

\begin{figure}
\hspace{2cm}\epsfig{file=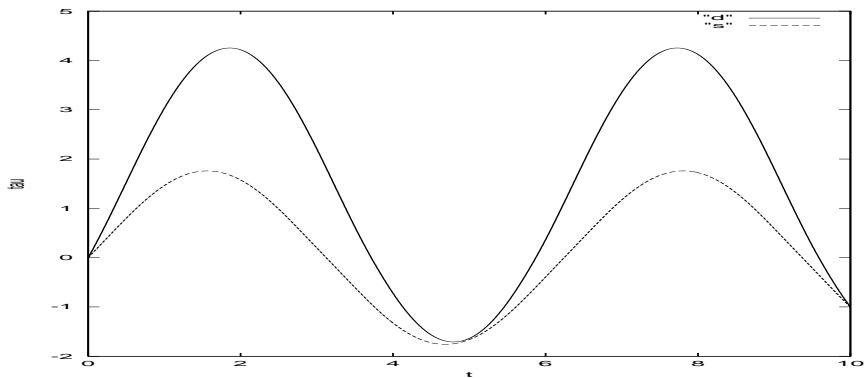,height=5cm,width=12cm,angle=0}
\vspace{.5cm}
\caption{Evolution of $\tau$ for a positive $\Lambda$ term in absence
of spinor mass. Only two solution are available (dust and stiff matter).}
\end{figure}

\begin{figure}
\hspace{2cm}\epsfig{file=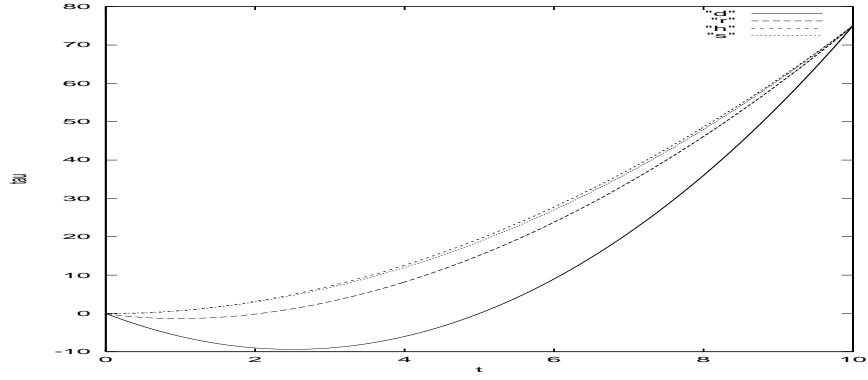,height=5cm,width=12cm,angle=0}
\vspace{.5cm}
\caption{Evolution of $\tau$ absence of $\Lambda$ term. 
For dust and radiation expansion starts at some positive value of $t$,
and the rate of expansion is rather rapid.}
\end{figure}

\begin{figure}
\hspace{2cm}\epsfig{file=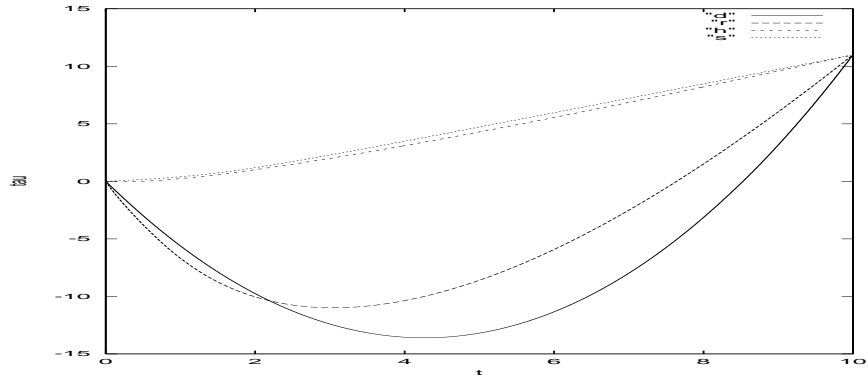,height=5cm,width=12cm,angle=0}
\vspace{.5cm}
\caption{Evolution of $\tau$ in absence of $\Lambda$  and massive terms. 
There is no isotropization in the case concerned.}
\end{figure}

\begin{figure}
\hspace{2cm}\epsfig{file=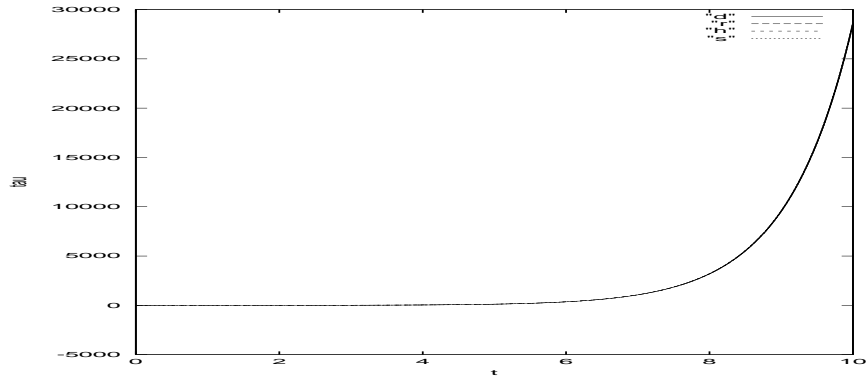,height=5cm,width=12cm,angle=0}
\vspace{.5cm}
\caption{Evolution of $\tau$ with a negative $\Lambda$. Effects of mass,
interaction and perfect fluid are completely suppressed by $\Lambda$-term.}
\end{figure}

\begin{figure}
\hspace{2cm}\epsfig{file=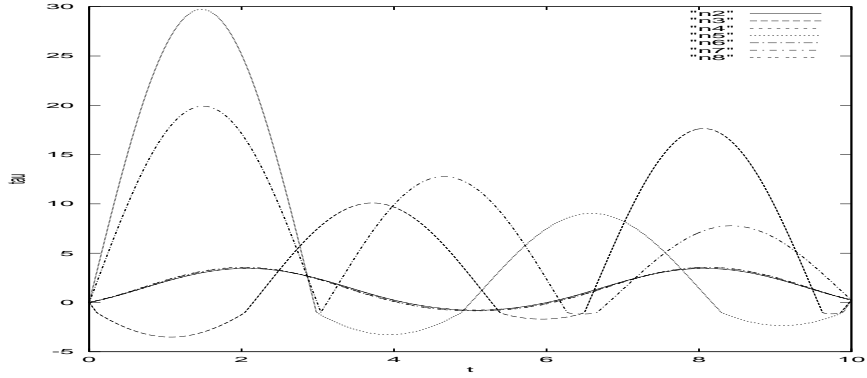,height=5cm,width=12cm,angle=0}
\vspace{.5cm}
\caption{Evolution of $\tau$ with a positive $\Lambda$ for different values
of $n$ (power of nonlinearity). As one sees, for $n$ even, the solutions 
more or less coincide, while it is not the case when $n$ is odd.}
\end{figure}

\begin{figure}
\hspace{2cm}\epsfig{file=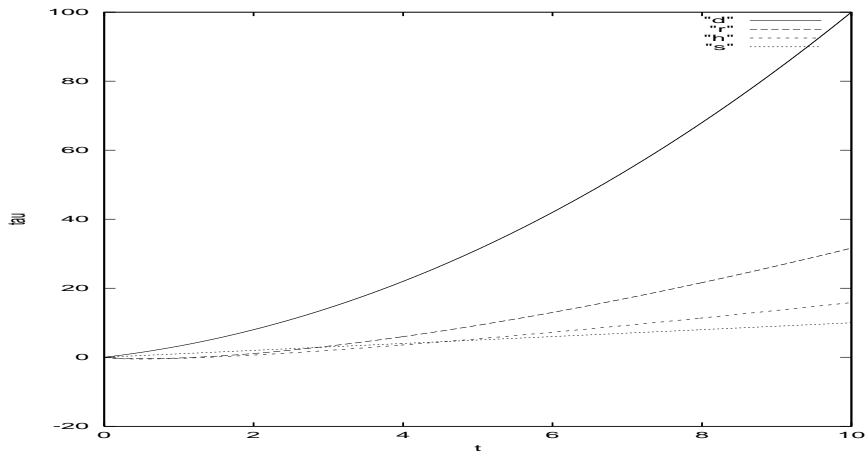,height=6cm,width=12cm,angle=0}
\vspace{.5cm}
\caption{Evolution of $\tau$ when BI universe is filled with
perfect fluid only.} 
\end{figure}


\section{Conclusion}

Within the framework of the simplest model of interacting spinor 
and scalar fields it is shown that the $\Lambda$ term plays very 
important role in BI cosmology. In particular, it invokes 
oscillations in the model which is not the case when $\Lambda$ 
term remains absent. It should be noted that regularity of the 
solutions obtained by virtue of $\Lambda$ term, specially for 
the linear spinor field does not violate dominant energy condition, 
while this is not the case when regular solutions are attained by 
means of nonlinear term. Note that in presence of $\Lambda$-term 
the role of other
parameters such as order of nonlinearity $n$, perfect fluid 
parameter $\zeta$ and spinor mass in the evolution process are 
rather local, while the global process are totally determined
by the $\Lambda$-term, e.g.,
for a positive $\Lambda$ we have always oscillatory mode, while for
a negative $\Lambda$ solution is always inflation-like.


\end{document}